\documentclass[twocolumn,showpacs,aps,prd]{revtex4}
\usepackage{graphicx}
\usepackage{dcolumn}
\usepackage{amsmath}
\usepackage{epsfig}
\input babarsym

\begin{document}

\title{\boldmath
First observation of the isospin violating decay $J/\psi\rightarrow
\Lambda\bar{\Sigma}^{0}+c.c.$  }
\author{
{\small M.~Ablikim$^{1}$, M.~N.~Achasov$^{5}$, D.~J.~Ambrose$^{39}$,
F.~F.~An$^{1}$, Q.~An$^{40}$, Z.~H.~An$^{1}$, J.~Z.~Bai$^{1}$,
Y.~Ban$^{27}$, J.~Becker$^{2}$, N.~Berger$^{1}$
% ...
\\
\vspace{0.2cm}
(BESIII Collaboration)\\
\vspace{0.2cm} {\it $^{1}$ Institute of High Energy Physics, Beijing 100049, P. R. China\\
$^{2}$ Bochum Ruhr-University, 44780 Bochum, Germany\\
% ...
$^{5}$ G.I. Budker Institute of Nuclear Physics SB RAS (BINP), Novosibirsk 630090, Russia\\
% ...
$^{27}$ Peking University, Beijing 100871, P. R. China\\
% ...
$^{39}$ University of Rochester, Rochester, New York 14627, USA\\
$^{40}$ University of Science and Technology of China, Hefei 230026, P. R. China\\
% ...
}} \vspace{0.4cm} }

\begin{abstract}

Using a sample of $(225.2\pm 2.8)\times 10^6$ $J/\psi$ events
collected with the BESIII detector, we present results of a study of
$J/\psi\rightarrow \gamma\Lambda\bar{\Lambda}$ and report the first
observation of the isospin violating decay
$J/\psi\rightarrow\Lambda\bar{\Sigma}^{0}+c.c.$, in which
$\bar{\Sigma}^{0}$ decays to $\gamma \bar{\Lambda}$. The measured
branching fractions are
$\mathcal{B}(J/\psi\rightarrow\bar{\Lambda}\Sigma^{0}$) =
$(1.46\pm0.11\pm0.12) \times10^{-5}$ and
$\mathcal{B}(J/\psi\rightarrow\Lambda\bar{\Sigma^{0}}$) =
$(1.37\pm0.12\pm0.11) \times10^{-5}$.  We search for $\Lambda(1520)
\rightarrow \gamma \Lambda$ decay, and find no evident signal, and an
upper limit for the product branching fraction
$\mathcal{B}(J/\psi\rightarrow\Lambda\bar{\Lambda}(1520)+c.c.)\times
\mathcal{B}(\Lambda(1520)\rightarrow\gamma\Lambda)<4.1 \times10^{-6}$
is set at the 90\% confidence level. We also report the observation of
$\eta_{c}\rightarrow\Lambda\bar{\Lambda}$ in $J/\psi \rightarrow
\gamma \eta_{c}$, $\eta_{c}\rightarrow\Lambda\bar{\Lambda}$ and
measure the branching fraction
$\mathcal{B}(\eta_{c}\rightarrow\Lambda\bar{\Lambda}) =(1.16\pm0.12
\mbox{(stat)}\pm0.19\mbox{(syst)}\pm0.28\mbox{(PDG)})\times10^{-3}$.

\end{abstract}


\maketitle

\section{Introduction}
\label{sec:intro}

 The study of charmonium meson decays into baryon pairs is an
important field that intersects particle and nuclear physics, and
provides a novel means for exploring various properties of
baryons~\cite{Tab1}.  The decay
$J/\psi\rightarrow\bar{\Lambda}\Sigma^{0}+c.c.$ is an isospin
symmetry breaking decay, and a measurement of its branching fraction
will help elucidate isospin-breaking mechanisms in $J/\psi\to
B_8\bar B_8$ decays~\cite{kopke,kowalski}. Until now, only an upper
limit on the branching fraction of ${\mathcal B}(
J/\psi\rightarrow\bar{\Lambda}\Sigma^{0}+c.c.)< 1.5\times10^{-4}$
has been set at the 90\% confidence level (C.L.) by the MarkI
Collaboration, based on a study of $J/\psi\rightarrow \gamma
\Lambda\bar{\Lambda}$ ~\cite{mark1}.

The electromagnetic decays of hyperons $\Lambda^*\to\gamma\Lambda$
provide clean probes for examining the internal structure of
$\Lambda^*$ hyperon resonances~\cite{dulat}.  For example,
predictions for the radiative decay $\Lambda(1520)\to \gamma
\Lambda$ have been made in a number of frameworks including: a
nonrelativistic quark model~\cite{kaxiras,darewych}; a relativistic
constituent quark model~\cite{warns}; the MIT bag
model~\cite{kaxiras}; the chiral bag model~\cite{umino}; an
algebraic model of hadron structure~\cite{bijker}; and a chiral
quark model~\cite{LangYu}.  In contrast, experimental measurements
have been sparse~\cite{mast, bertini, antipov, taylor}. The
radiative decays $\Lambda^*\to\gamma\Lambda$ can be studied with
$J/\psi \to \gamma\Lambda\bar{\Lambda}$ events.

The $J/\psi \to \gamma\Lambda\bar{\Lambda}$ events can also originate
from radiative $J/\psi \to \gamma \eta_c$ decays followed by $\eta_c$
decays to $\Lambda\bar\Lambda$. To date, $\eta_c \to
\Lambda\bar\Lambda$ has only been observed in $B^{\pm}\to
\Lambda\bar\Lambda K^{\pm}$ decays by the Belle
experiment~\cite{belle}.  A measurement of $\eta_c \to
\Lambda\bar\Lambda$ in $J/\psi$ radiative decays provides useful
information in addition to Belle's measurement in $B$ decays.

In this paper,  we report the first observation of the isospin
violating decay $J/\psi\rightarrow\Lambda\bar{\Sigma}^{0}+c.c.$, a
new measurement of the branching fraction for $\eta_c \to \Lambda \bar
\Lambda$ and the results of a search for the radiative decay $\Lambda(1520) \to
\gamma \Lambda$.

\section{Detector and Monte Carlo simulations}
\label{sec:det}

The analysis is based on analyses of
 $J/\psi \to \gamma \Lambda \bar\Lambda$ events contained in
a sample of $(225.2\pm 2.8)\times 10^6$ $J/\psi$ events~\cite{number}
accumulated with the Beijing Spectrometer III (BESIII) operating at the Beijing
Electron-Position Collider II (BEPCII)~\cite{dect8}.

BEPCII is a double ring $e^+e^-$ collider with a design peak
luminosity of $10^{33}\rm{cm}^{-2}\rm{s}^{-1}$ with beam currents of
0.93 A. The BESIII detector consists of a cylindrical core comprised
of a helium-based main drift chamber (MDC), a plastic scintillator
time-of-flight (TOF) system, and a CsI(Tl) electromagnetic
calorimeter (EMC) that are all enclosed in a superconducting
solenoidal magnet that provides a 1.0 T axial magnetic field. The
solenoid is supported by an octagonal flux-return yoke that contains
resistive-plate-chamber muon-identifier modules interleaved with
plates of steel. The acceptance for charged particles and photons is
93\% of 4$\pi$~sr, and the charged-particle momentum and photon
energy resolutions at 1 GeV are 0.5\% and 2.5\%, respectively.

The responses of the BESIII detector are modeled with a Monte Carlo
(MC) simulation based on {\sc geant4}~\cite{geant1,geant2}. {\sc
evtgen}~\cite{evtgen} is used to generate $J/\psi \to \Lambda
\bar{\Sigma}^0 +c.c.$ events with an angular distribution of $1+
\alpha \cos^2 \theta$, where $\theta$ is the polar angle of the
baryon in the $J/\psi$ rest frame and $\alpha$ is a parameter
extracted in fits to data described below.  The $J/\psi\to\gamma
\eta_c$ decays are generated with an angular distribution of
$1+\cos^2\theta_\gamma$ and a phase-space distribution for $\eta_c
\to \Lambda\bar{\Lambda}$, and effect of spin-correlation is not
considered in the MC simulation for $\eta_c \to
\Lambda\bar{\Lambda}$ decay. Inclusive $J/\psi$ decays are produced
by the MC event generator {\sc kkmc}~\cite{kkmc}, the known $J/\psi$
decay modes are generated by {\sc evtgen}~\cite{evtgen} with
branching fractions set at their Particle Data Group (PDG) world
average values~\cite{pdg}, and the remaining unknown decays are
generated with {\sc lundcharm}~\cite{lund}.

\section{Data Analysis}
\label{sec:select}

Charged tracks in the BESIII detector are reconstructed from
track-induced signals in the MDC. We select tracks  within $\pm20$
cm of the interaction point in the beam direction and within 10 cm
in the plane perpendicular to the beam; the track directions are
required to be within the MDC fiducial volume, $|\cos\theta| <
0.93$. Candidate events are required to have four charged tracks
with net charge zero. The $\Lambda \bar\Lambda$ pair is
reconstructed using the $\Lambda \to p \pi^-$, and $\bar \Lambda \to
\bar p \pi^+$ decay modes. We loop over all the combinations of
positive and negative charged track pairs and require that at least
one $(p\pi^-)(\bar p \pi^+)$ track hypothesis successfully passes
the $\Lambda$ and $\bar \Lambda$'s vertex finding algorithm.

If there is more than one accepted
$(p\pi^-)(\bar p \pi^+)$ combination in an event, the candidate with
minimum value of $(M_{p\pi^-} - M_{\Lambda})^2 + (M_{\bar p\pi^+} -
M_{\bar \Lambda})^2$ is selected, where $M_{p\pi^-}$ ($M_{\bar
p\pi^+}$) and $M_{\Lambda}$ ($M_{\bar\Lambda}$) are the measured
mass and its expected value. Since there are differences in the detection
efficiencies between data and the MC simulation for low-momentum
proton and antiprotons~\cite{ppbar}, we reject events containing any
proton or antiproton track candidate with momentum below 0.3 GeV/$c$.

Electromagnetic showers are reconstructed from clusters of energy
deposits in the EMC. The energy deposited in nearby TOF counters is
added to improve the reconstruction efficiency and energy
resolution. Showers identified as photon candidates are required to
satisfy fiducial and shower-quality requirements: {\it e.g.},
showers in the barrel region ($|\cos\theta|<0.80$) must have a
minimum energy of 25 MeV, while those from end caps
($0.86<|\cos\theta|<0.92$) must have at least 50 MeV.  To suppress
showers generated by charged particles, we require that the photon
candidate direction is at least $5^{\circ}$ away from its nearest
proton and charged pion tracks, and at least $30^{\circ}$ away from
the nearest antiproton track, since more EMC showers tend to be
found near the direction of the antiproton. This requirement
decreases the signal efficiency by 18\% for $J/\psi\to \Lambda
\bar{\Sigma}^0 (\bar{\Sigma}^0\to \gamma \bar{\Lambda})$ compared to
that for $ J/\psi\to \bar\Lambda \Sigma^0 (\Sigma^0\to \gamma
\Lambda)$ since the photon from the radiative $\bar{\Sigma}^0\to
\gamma \bar{\Lambda}$ decay is closer to the direction of the
antiproton. Requirements on the EMC cluster timing are used to
suppress electronic noise and energy deposits that are unrelated to
the event. A four-constraint (4C) energy-momentum conservation
kinematic fit is performed to the $\gamma \Lambda \bar\Lambda$
hypothesis. For events with more than one photon candidate, the
combination with the minimum $\chi^2_{4C}$ is selected. In addition,
we also require $\chi^{2}_{4C}<45$ in order to suppress backgrounds
from the decays $J/\psi \rightarrow \Lambda \bar\Lambda$,  $\Sigma^0
\bar{\Sigma}^0$ and $\Lambda\bar\Lambda \pi^0$.
\begin{figure}[tb]
\centering
\includegraphics[width=6.0cm]{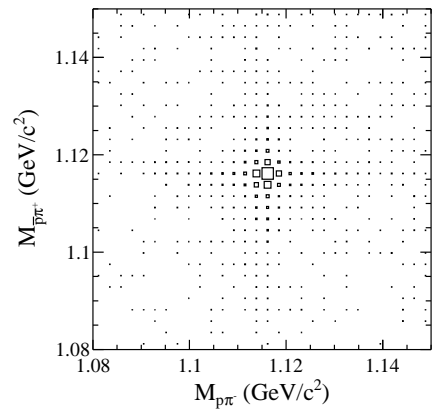}
\caption{A scatter plot of $M_{p\pi^{-}}$ {\it versus}
$M_{\bar{p}\pi^{+}}$ for selected candidate events.}
\label{scatter_R}
\end{figure}

\begin{figure}
\centering
\includegraphics[width=4.2cm]{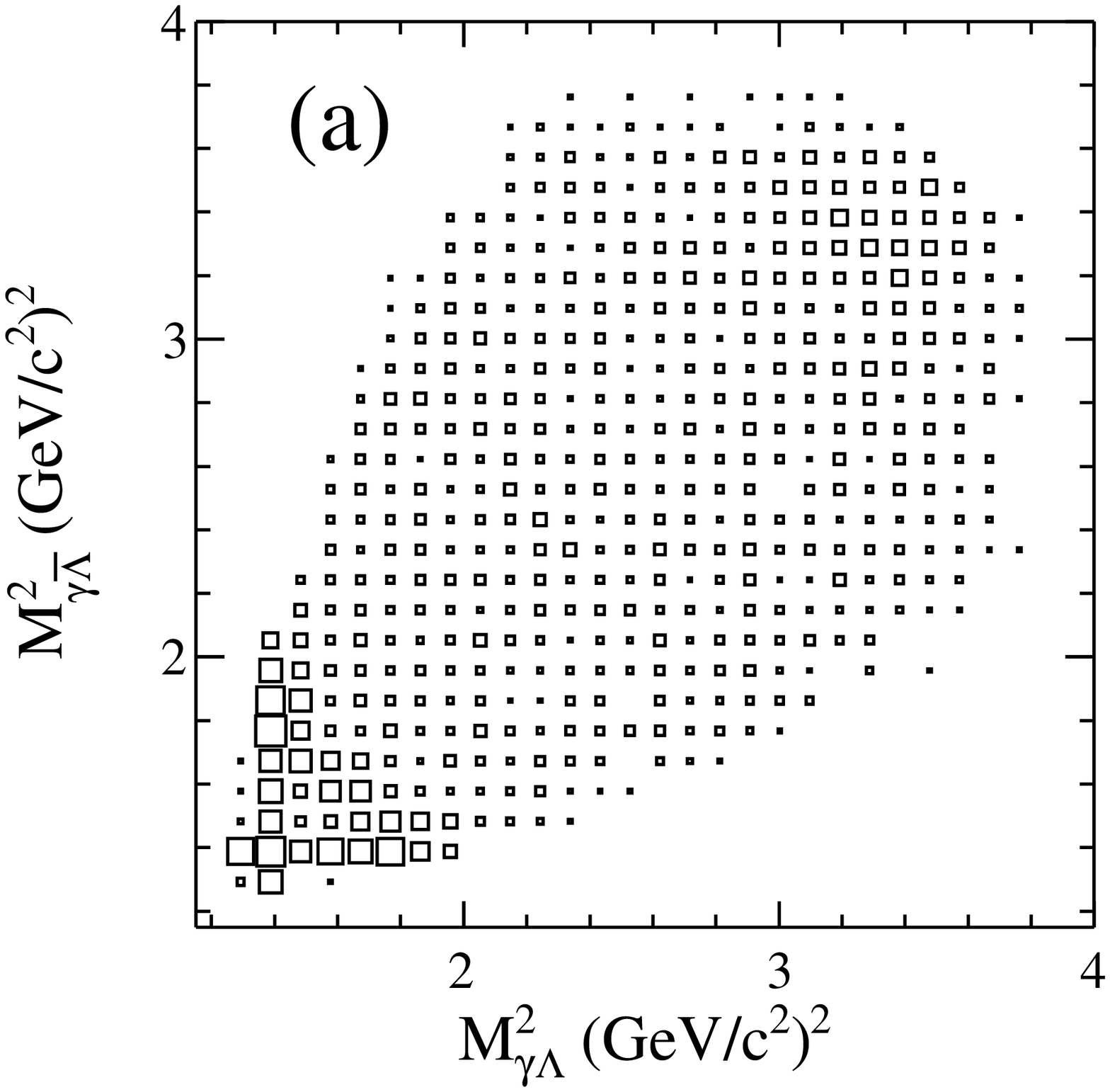}
\includegraphics[width=4.2cm]{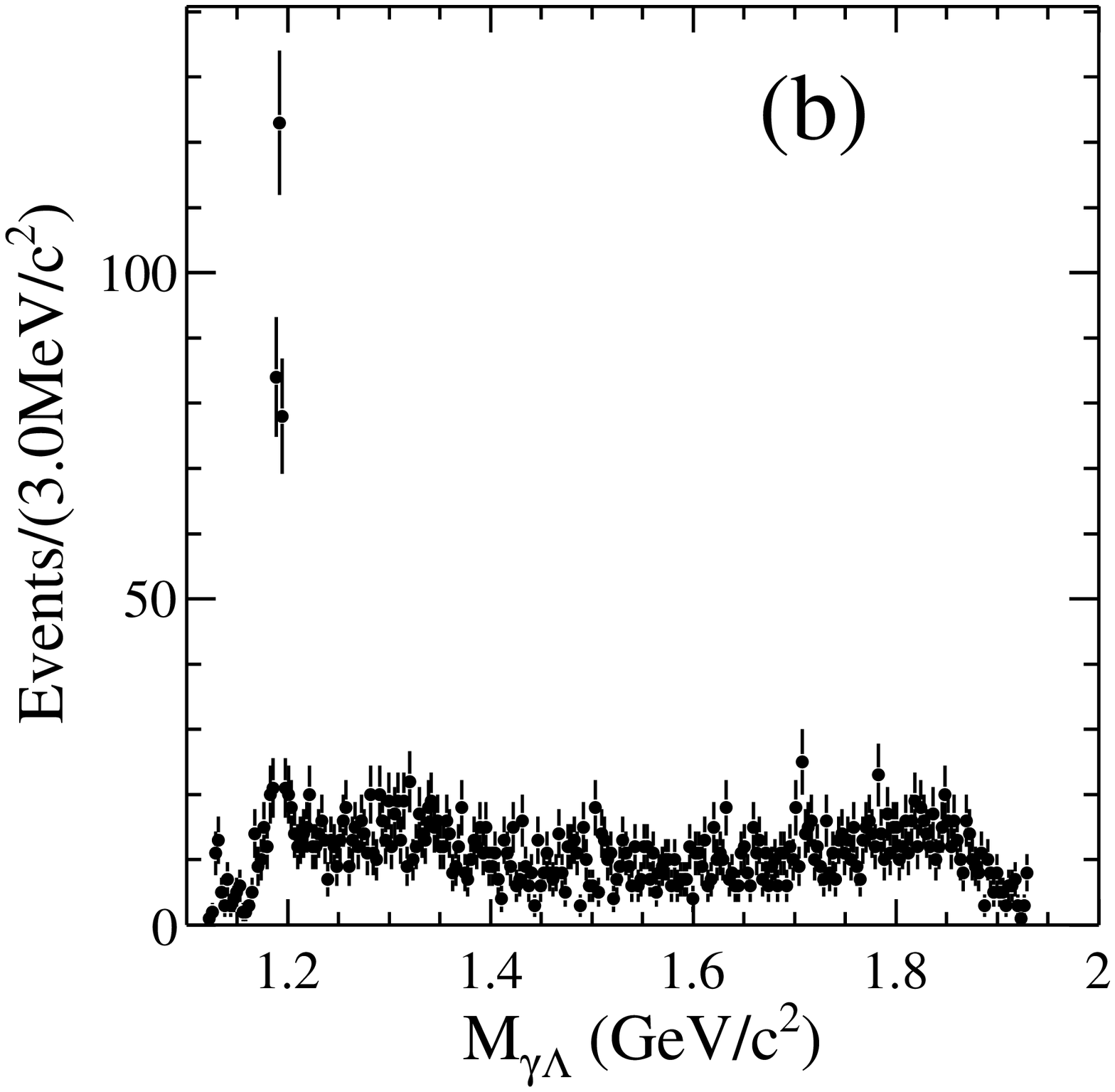}
\includegraphics[width=4.2cm]{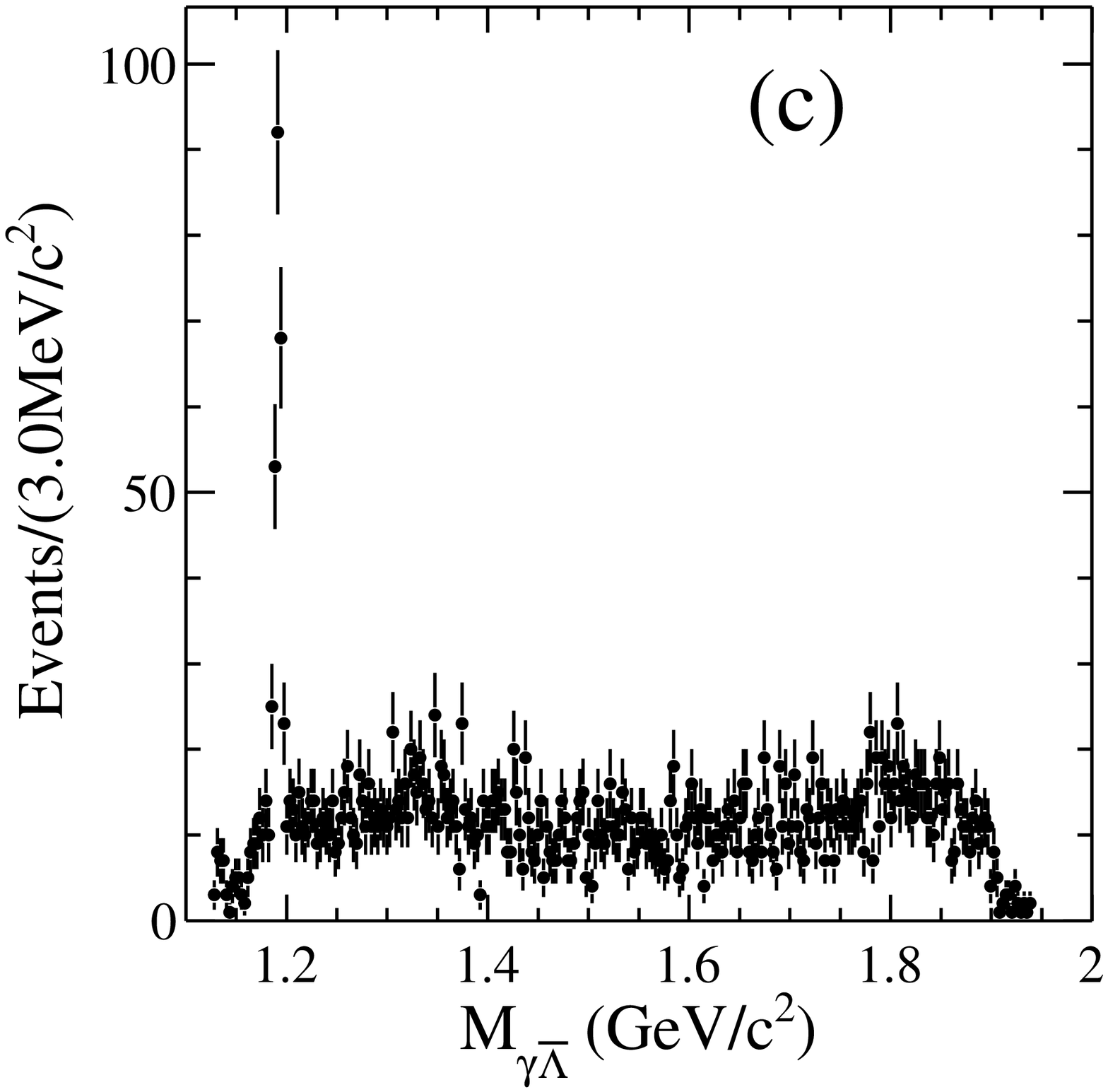}
\includegraphics[width=4.2cm]{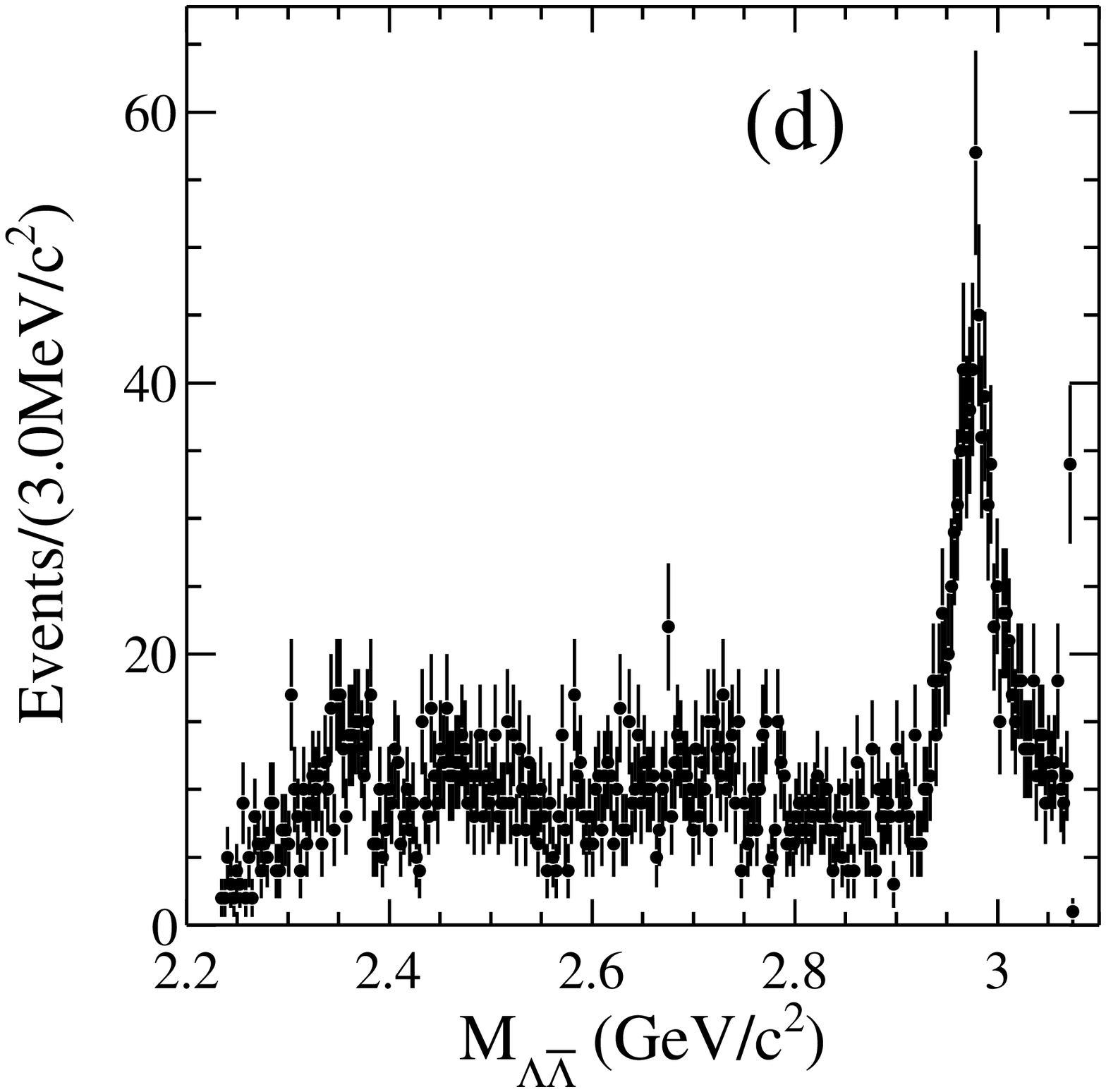}
\caption{ (a)  An $M^2(\gamma\bar{\Lambda})$ (vertical) {\it versus}
$M^2(\gamma\Lambda)$ (horizontal) Dalitz plot for selected events
and the (b) $\gamma\Lambda$ , (c) $\gamma\bar{\Lambda}$ and (d)
$\Lambda\bar{\Lambda}$ invariant mass distributions for the selected
$J/\psi\rightarrow\gamma\Lambda\bar{\Lambda}$ event sample.}
\label{plot_4C} 
\end{figure}

A scatter plot of $M_{p \pi^{-}}$ {\it versus} $M_{\bar{p} \pi^{+}}$
for events that survive the above requirements is shown in
Fig.~\ref{scatter_R}), where a cluster of $\Lambda$ and
$\bar\Lambda$ signals is evident.  To select $J/\psi \rightarrow
\gamma \Lambda\bar\Lambda$ signal events, we require $|M_{p\pi^-} -
M_{\Lambda}|<5$ MeV/$c^2$ and $|M_{\bar p\pi^+} - M_{\bar
\Lambda}|<5$ MeV/$c^2$.
 An $M^2(\gamma\bar{\Lambda})$ (vertical)
{\it versus} $M^2(\gamma\Lambda)$ (horizontal) Dalitz plot for these
events is shown in Fig.~\ref{plot_4C} (a); the $\gamma \Lambda$ and
$\gamma \bar\Lambda$ mass spectra are shown in Fig.~\ref{plot_4C}
(b) and (c). Prominent signals of the $\Sigma^0$ and
$\bar{\Sigma}^0$, corresponding to
$J/\psi\rightarrow\Lambda\bar{\Sigma}^{0}+c.c.$ decays, are
observed. On the other hand, no obvious signal for $\Lambda(1520)\to
\gamma \Lambda$ is seen.  A clear $\eta_c$ signal can be seen in the
$\Lambda\bar\Lambda$ mass spectrum shown in Fig.~\ref{plot_4C} (d),
while no significant enhancement at other $\Lambda\bar\Lambda$
masses is evident.

\begin{figure}
\centering
\includegraphics[width=5.2cm]{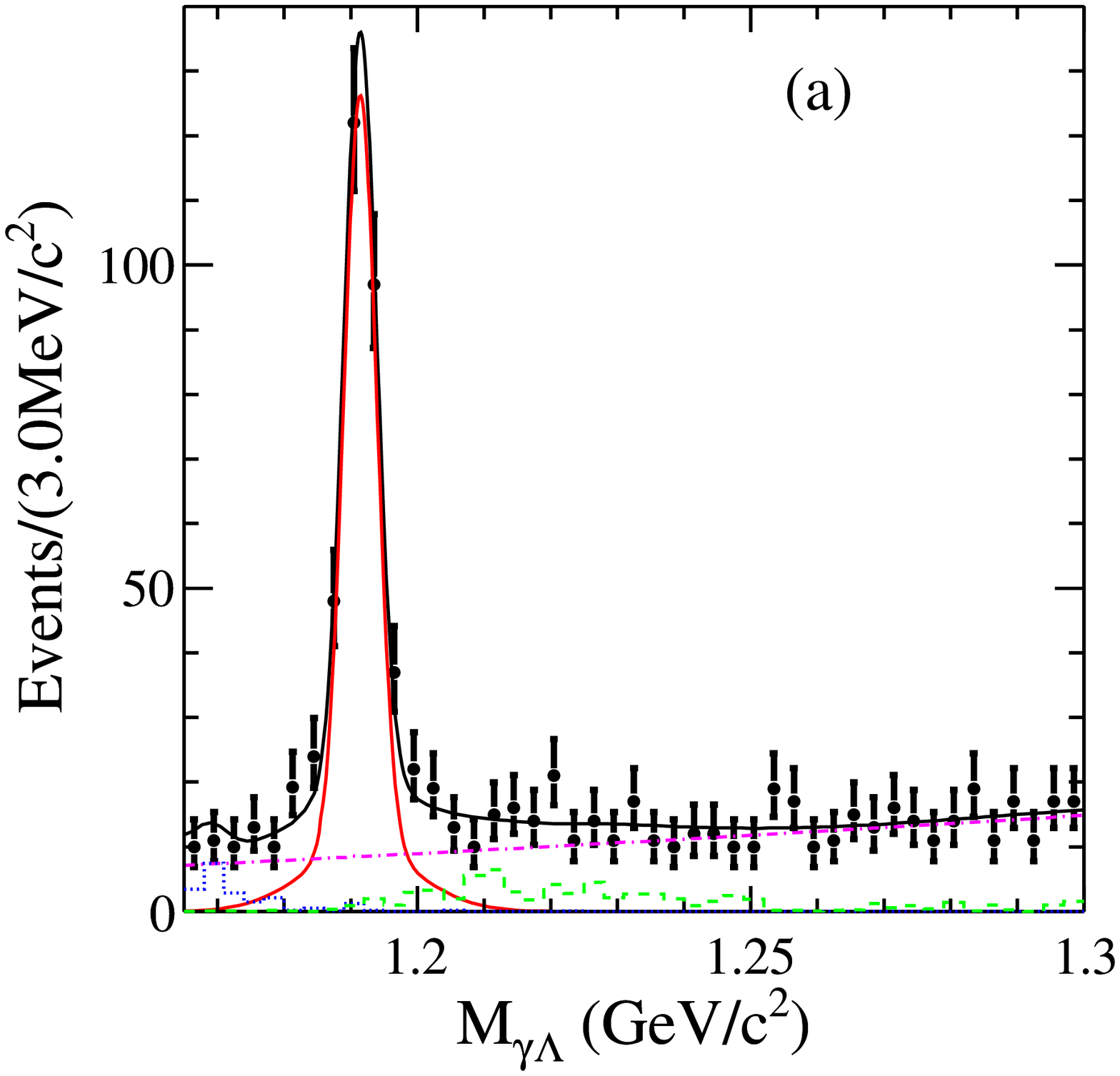}
\includegraphics[width=5.2cm]{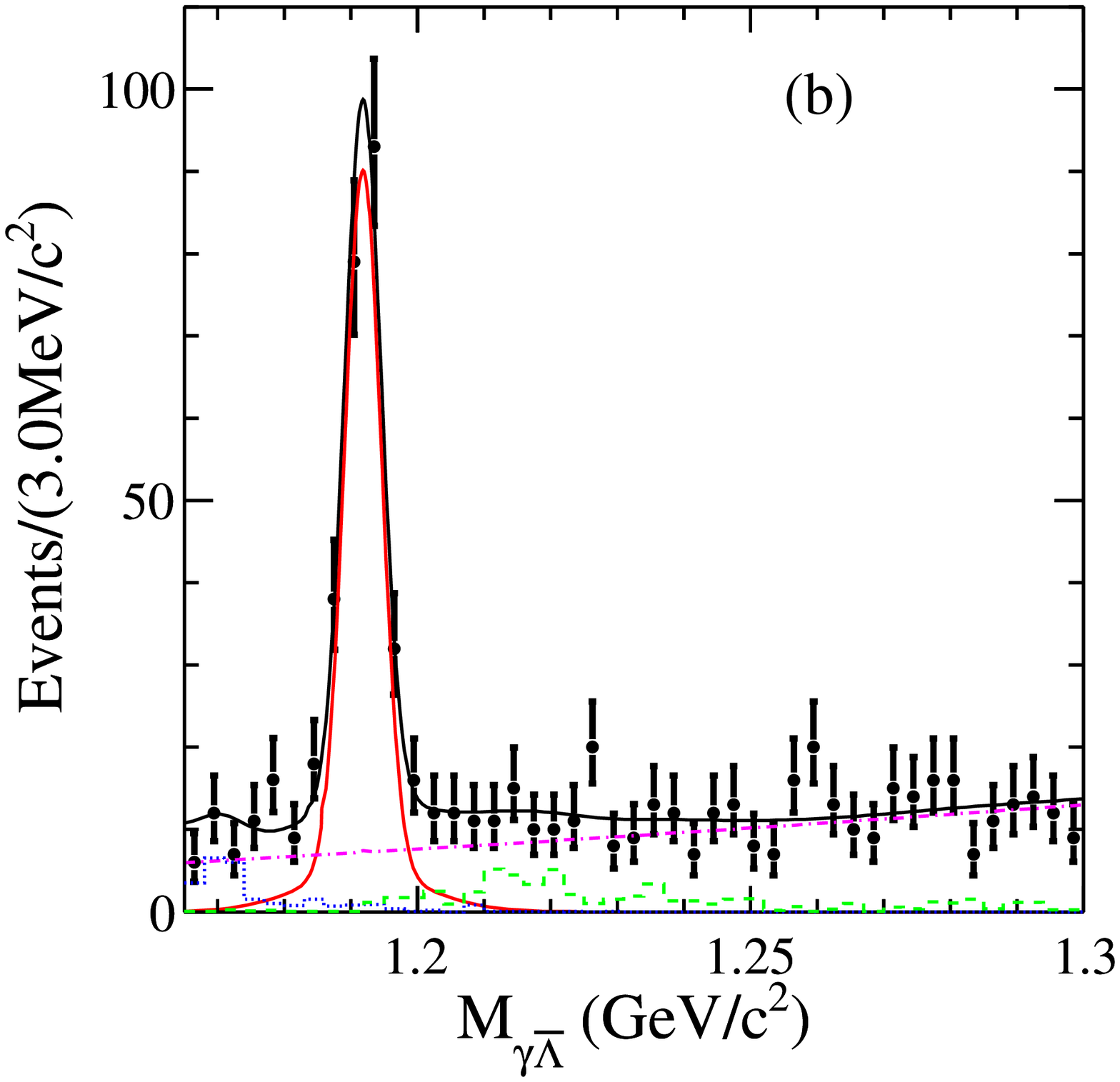}
\caption{The results of the fit for the $\Sigma^{0}$ (a) and
$\bar{\Sigma^{0}}$ (b). The points with error bars are data. The fit
results are shown by the black solid curves. The light (red) solid
curves are the signal shapes. The (blue) dotted histograms are from
the normalized $J/\psi\rightarrow\Lambda\bar{\Lambda}$ background;
the (green) dashed histograms are from the normalized
$J\psi\rightarrow\Sigma^{0}\bar{\Sigma^{0}}$ background. The
(magenta) dot-dashed curves show the nonresonant (phase-space)
background polynomial.} \label{sigma0_fit_R}
\end{figure}

For the $J/\psi \rightarrow \Lambda \bar{\Sigma}^0 +c.c.$ study, we
apply the same requirements to a sample of $225\times 10^6$
MC-simulated inclusive $J/\psi$ events and find that the primary
backgrounds come from  $J/\psi \rightarrow \Lambda \bar\Lambda$,
$\Sigma^0 \bar{\Sigma}^0$ and $\Lambda\bar\Lambda \pi^0$  decays,
where either a cluster in the EMC unrelated to the event is
misidentified as a photon candidate or one of the photons from the
$\Sigma^{0}\bar{\Sigma}^{0}$ or $\pi^0$ decay is undetected in the
EMC.  Normalized $M(\gamma\Lambda)$ and $M(\gamma\bar{\Lambda})$
distributions from the events that survive the application of the 4C
kinematic fit, shown as dotted and dashed histograms in
Figs.~\ref{sigma0_fit_R} (a) and (b), show no sign of peaking in the
$\Sigma^0$ or $\bar{\Sigma}^0$ mass regions. Another potential
source of background is from $J/\psi \rightarrow \gamma \eta_c$
($\eta_c \to\Lambda\bar\Lambda$) decay and nonresonant $J/\psi \to
\gamma \Lambda\bar\Lambda$, which contribute a smooth background
under the signal region, shown as dot-dashed curves in
Figs.~\ref{sigma0_fit_R} (a) and (b). The expected backgrounds are
$105\pm10$ ($95\pm9$) events in the $\Sigma^0$ ($\bar{\Sigma}^0$)
signal region for $J/\psi \to \bar{\Lambda}\Sigma^0$ ($J/\psi \to
\Lambda\bar{\Sigma}^0$) as listed in Table~\ref{sum:eff_yield}. The
signal region is defined as being within $\pm 3\sigma$ of the
nominal $\Sigma^0$ ($\bar{\Sigma}^0$) mass. It should be noted that
the background events from the nonresonant $J/\psi \rightarrow
\gamma \Lambda\bar{\Lambda}$ are not counted and are accounted for
by the floating polynomial function discussed below.

Unbinned maximum likelihood (ML) fits are used  to determine the
$\Lambda \bar{\Sigma}^0$ and $\bar{\Lambda} {\Sigma}^0$ event
yields. The signal probability density function (PDF) for $\Sigma^0$
($\bar{\Sigma}^0$) from $J/\psi \to \bar\Lambda\Sigma$
($\Lambda\bar{\Sigma}^0$) is represented by a double-Gaussian
function with parameters determined from the MC simulation except
for the Gaussian widths, which are allowed to float. Backgrounds
from $J/\psi \rightarrow \Lambda \bar\Lambda$ and $\Sigma^0
\bar{\Sigma}^0$ are fixed to their MC simulations at their expected
intensities.  The remaining background is described by a
second-order polynomial function with parameters that are allowed to
float. The fitting ranges for both the $\Sigma^0$ and the
$\bar{\Sigma}^0$ are $1.165 - 1.30$ GeV/$c^2$.
Figures~\ref{sigma0_fit_R} (a) and (b) show the results of the fits
to $\Sigma^0$ and $\bar{\Sigma}^0$. The fitted yields are $308\pm24$
and $234\pm21$ signal events for $J/\psi \to \bar{\Lambda}\Sigma^0$
and $\Lambda\bar{\Sigma}^0$, respectively. The goodness of fit is
estimated by using a $\chi^2$ test method with the data
distributions regrouped to ensure that each bin contains more than
10 events. The test gives $\chi^2/n.d.f= 28.1/37 = 0.76$ for $J/\psi
\to \bar{\Lambda}\Sigma^0$ and $\chi^2/n.d.f= 43.5/37 = 1.2$ for
$J/\psi \to \Lambda\bar{\Sigma}^0$, where $n.d.f.$ is the number of
degrees of freedom.

\begin{figure}
\centering
\includegraphics[width=6.0cm]{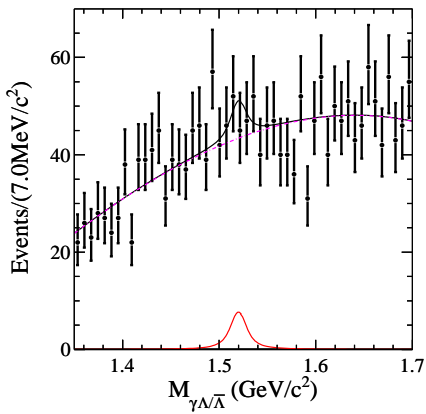}
\caption{The results of the fit for the $\Lambda(1520)$. The points
with error bars are data. The fit result is shown by the black solid
curve; the (magenta) dashed curve is the background polynomial and
the (red) light solid curve is the $\Lambda(1520)$ signal shape.
[Here the $M(\gamma\Lambda)$ and $M(\gamma\bar{\Lambda})$ mass
distributions are combined.]  }
 \label{1520_cl}
\end{figure}

In the higher $\gamma \Lambda$ ($\gamma \bar{\Lambda})$ invariant
mass regions, shown in Figs.~\ref{plot_4C} (b) and (c), no obvious
signals for $\Lambda(1520)\rightarrow \gamma \Lambda$
($\bar{\Lambda}(1520)\rightarrow \gamma \bar{\Lambda}$) are evident.
We require that the invariant mass of $\Lambda \bar\Lambda$ is less
than 2.9 GeV/$c^2$ to further suppress combinatorial backgrounds
from $J/\psi \rightarrow \Lambda \bar\Lambda$, $\Sigma^0
\bar{\Sigma}^0$, $\Lambda\bar\Lambda \pi^0$ and $J/\psi \to \gamma
\eta_c (\eta_c \to \Lambda\bar\Lambda)$ decays. After the above
requirement, only $14\pm1$ events from these background decay modes
remain. In the surviving combined $M(\gamma\Lambda)$ and
$M(\gamma\bar{\Lambda})$ mass spectrum, shown in Fig.~\ref{1520_cl},
there is no evidence for a $\Lambda(1520)$ signal above expectations
for a phase-space distribution of $J/\psi \to \gamma
\Lambda\bar\Lambda$.

In the ML fit to the Fig.~\ref{1520_cl} distribution, the
$\Lambda(1520)$ signal PDF is represented by a Breit-Wigner (BW)
function convolved with a double-Gaussian resolution function, with
parameters determined from the fit to the $\Sigma^0$ data. The shape
for the nonresonant background is described by a second-order
polynomial function, and the background yield and its PDF parameters
are allowed to float in the fit. The mass range used for the
$\Lambda(1520)$ fit is $1.35 - 1.70$ GeV/$c^2$. Figure~\ref{1520_cl}
shows the result of the fit to $\Lambda(1520)$, which returns a
$\Lambda(1520)$ signal yield of $31\pm24$ events. The goodness of
fit is $\chi^2/n.d.f= 45.9/45 =1.02$. Using a Bayesian method, an
upper limit for the number of $\Lambda(1520)$ signal events is
determined to be 62.5 at the 90\% confidence level (C.L.). The
signal yields and the efficiencies for the analyses of $J/\psi \to
\bar{\Lambda}\Sigma^0$ ($\Lambda\bar{\Sigma}^0$) and
$\Lambda\bar{\Lambda}(1520)+c.c.$ are summarized in
Table~\ref{sum:eff_yield}.

For the $J/\psi \to \gamma \eta_c (\eta_c \to \Lambda\bar\Lambda)$
analysis, the dominant backgrounds remaining after event selection
are from $J/\psi \rightarrow \Sigma^0 \bar{\Sigma}^0$ and
$\Lambda\bar{\Sigma}^0 +c.c.$.
 The expected number of events in the signal region
 from these two sources is $637\pm52$,
 as listed in Table~\ref{sum:eff_yield}.
 These backgrounds are incoherent ({\it i.e.,} do not interfere
 with the $\eta_c$ signal amplitude).
In addition, there is an irreducible background from nonresonant
 $J/\psi\to \gamma \Lambda\bar\Lambda$ decays that is potentially
  coherent with the signal process
 ({\it i.e.,} may interfere with the $\eta_c$ signal
 amplitude).

\begin{figure}[htbp]
\begin{center}
\includegraphics[width=6.0cm]{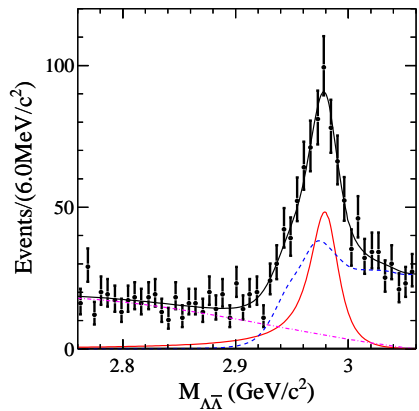}
\caption{ The $\eta_c$ mass distribution and fit results. Points
with error bars are data. The fit result is shown as a black solid
curve, the (red) light solid curve is the signal shape, the (blue)
dashed curve is the combined incoherent background from the
$J/\psi\rightarrow\Sigma^{0}\bar{\Sigma^{0}}$,
$\Lambda\bar{\Sigma^{0}}+c.c.$, the (magenta) dot-dashed-curve is
the nonresonant background. }
 \label{etac_bgnew}
\end{center}
\end{figure}
\begin{table}[hbtp]
  \footnotesize
 \centering \caption{For each decay mode, the number of signal events ($N_S$),
   the number of expected background events ($N_B$) in the signal region
   (nonresonant $J/\psi \rightarrow \gamma \Lambda\bar{\Lambda}$  background is excluded),
   and the MC efficiency
  ($\varepsilon$) for signal are given. The error on $N_S$ is
  statistical only, and the signal regions are defined to be within
  $\pm3\sigma$ of the nominal $\Sigma^0$ and
  $\Lambda(1520)$ masses.}
  \label{sum:eff_yield}
     \renewcommand{\arraystretch}{1.1}
     \begin{tabular}{l|ccc}
        \hline
        \hline

        Modes                         & $N_S$    ~~ & ~~$N_B$  &
        ~~$\epsilon$(\%)\\

        \hline
        $ J/\psi\to \bar\Lambda \Sigma^0 (\Sigma^0\to \gamma \Lambda)$   & $308\pm24$ &  $105\pm10$ & 21.7  \\

        $J/\psi\to \Lambda \bar{\Sigma}^0 (\bar{\Sigma}^0\to \gamma \bar{\Lambda})$              & $234\pm21$   & $95\pm9$ & 17.6  \\

        $ J/\psi\to \Lambda \bar{\Lambda}(1520) +c.c. (\bar{\Lambda}(1520)\to \gamma \bar{\Lambda})$              & $31\pm24$  & $14\pm1$  &  18.8  \\

        $J/\psi \to \gamma \eta_c (\eta_c\to \Lambda\bar\Lambda)$   & $360\pm38$    & $637\pm52$ & 19.8  \\

        \hline\hline
      \end{tabular}
\end{table}

  For the $\eta_c$ fit, the combined incoherent background
 is fixed to the shape and level of the MC simulation. The PDF
 for the coherent nonresonant background is described by a second-order
  polynomial, with yield and shape parameters that are floated
in the fit. For the lineshape for $\eta_c$ mesons produced via the
M1 transition, we use $(E_{\gamma}^3 \times \mbox{BW}(m)\times
\mbox{damping}(E_\gamma)) \otimes \mbox{Gauss}(0,\sigma) $, where
$m$ is the  $\Lambda\bar{\Lambda}$ invariant mass, $E_\gamma =
\frac{M^2_{J/\psi}-m^2}{2M_{J/\psi}}$ is the energy of the
transition photon in the rest frame of $J/\psi$,
$\mbox{damping}(E_\gamma)$ is a function that damps the divergent
low-mass tail produced by the $E^3_\gamma$ factor,  and
$\mbox{Gauss}(0,\sigma)$ is a Gaussian function that describes the
detector resolution.  The damping function used by the
KEDR~\cite{anashin} collaboration for a related process has the form
\begin{eqnarray}
\frac{E^{2}_{0}}{E_{0}E_{\gamma}+(E_{0}-E_{\gamma})^{2}},
\end{eqnarray}
where $E_{0}=\frac{M^2_{J/\psi}-M^2_{\eta_{c}}}{2M_{J/\psi}}$ is the
peak energy of the transition photon. On the other hand, the CLEO
experiment damped the $E^3_{\gamma}$ term by a factor
exp(-$E^2_{\gamma}/8\beta^2$), with $\beta = 65$ MeV~\cite{cleoc},
to account for the difference in overlap of the ground state wave
functions. We use the KEDR function in our default fit and use the
CLEO function as an alternative.  The difference between the results
obtained with the two damping functions is considered as a
systematic error associated with uncertainties in the line shape. In
the fit, the mass and width of $\eta_c$ are fixed to the recent
BESIII measurements: $M(\eta_c)=2984.3\pm0.8$ MeV/$c^2$ and
$\Gamma(\eta_c)= 32.0\pm1.6$ MeV~\cite{bes3etac}, and interference
between the nonresonant background and the $\eta_c$ resonance
amplitude is neglected~\cite{etacfit}. The mass range used for the
$\eta_c$ fit is $2.76 - 3.06$ GeV/$c^2$. Figure~\ref{etac_bgnew}
shows the result of the fit to $\eta_c$, which yields $(360\pm38)$
signal events. The goodness of the fit is $\chi^2/n.d.f= 42.7/43
=0.99$. The signal yield and efficiency are summarized in
Table~\ref{sum:eff_yield}.

\section{Systematic Uncertainties}
\label{sec:sys}

The systematic uncertainties on the branching fraction measurements
are summarized in Table~\ref{tab:error}. The systematic uncertainty
due to the charged tracking efficiency has been studied with control
samples of $J/\psi\rightarrow pK^{-}\bar{\Lambda}+c.c.$ and
$J/\psi\rightarrow \Lambda\bar{\Lambda}$ decays. The difference in
the charged tracking efficiency between data and the MC simulation
is 1\% per track.  The uncertainty due to the $\Lambda$ and
$\bar{\Lambda}$ vertex fit is determined to be 1\% for each
$\Lambda$ by using the same control samples.  The uncertainty due to
the photon reconstruction is determined to be 1\% for each
photon~\cite{number}.  The uncertainties due to the kinematic fit
are determined by comparing the efficiency as a function of
$\chi^2_{4C}$ value for the MC samples and the control samples of
$J/\psi\rightarrow\Lambda\bar{\Lambda}$ and
$J/\psi\rightarrow\Sigma^{0}\bar{\Sigma}^{0}$ events, in which zero
and two photons are involved in the final states. The differences in
the efficiencies between data and MC simulation are 2.1\% and 2.3\%
from the studies of $J/\psi\rightarrow\Lambda\bar{\Lambda}$ and
$J/\psi\rightarrow\Sigma^{0}\bar{\Sigma}^{0}$ events, respectively;
 we use $2.3\%$ as the systematic error due to the kinematic fit.

 The signal shape for the $\Sigma^0$ ($\bar{\Sigma}^0$) is described by a
 double-Gaussian function and the widths are floated in the nominal fit. An
alternative fit is performed by fixing the signal shape to the MC
simulation, and the systematic uncertainty is set based on the
change observed in the yield. In the fit to $\Lambda(1520)$, since
the shape of the signal is obtained from MC simulation, the
uncertainty is estimated by changing the mass and width of
$\Lambda(1520)$ by 1 standard deviation from their PDG world average
values~\cite{pdg}. This systematic error is determined in this way
to be 4.8\%.

In the $\eta_c$ fit, the mass resolution is fixed to the MC
simulation; the level of possible discrepancy is determined with a
smearing Gaussian, for which a nonzero $\sigma$ would represent a
MC-data difference in the mass resolution. The uncertainty
associated with a difference determined in this way is 1.1\%.
Changes in the mass and width of the $\eta_c$ used in the fit by 1
standard deviation from the recently measured BESIII
values~\cite{bes3etac}, produce a relative change in the signal
yield of 6.4\%. As mentioned above, damping functions from the KEDR
and CLEO collaborations were used in the fit to suppress the lower
mass tail produced by the $E^3_{\gamma}$ factor; the relative
difference in the yields between the two fits is 3.9\%.  The 7.6\%
quadrature sum of these uncertainties is used as the systematical
error associated with uncertainties in $\eta_c$ signal line-shape.

For the measurement of the $J/\psi \to \bar\Lambda \Sigma^0$
($\Lambda\bar{\Sigma}^0$), the expected number of background events
from the decays of $J/\psi \to \Lambda \bar\Lambda$ and
$\Sigma^0\bar{\Sigma}^0$ is fixed in the fit. To estimate the
associated uncertainty, we vary the number of expected background
events by 1 standard deviation from the PDG branching fraction
values~\cite{pdg}, which gives an uncertainty of 0.6\% (0.4\%) for
the $J/\psi \to \bar\Lambda \Sigma^0$ ($\Lambda\bar{\Sigma}^0$). In
the ML fit to $\eta_c$, the incoherent backgrounds from $J/\psi \to
\Sigma^0\bar{\Sigma}^0$ and $\bar\Lambda \Sigma^0 +c.c.$ are also
fixed at their expected numbers of events.  The uncertainty
associated with this is determined by changing the number of
expected incoherent background events by 1 standard deviation of the
PDG branching fraction values~\cite{pdg} for the $J/\psi \to
\Sigma^0\bar{\Sigma}^0$ and the measured value for $J/\psi \to
\bar\Lambda \Sigma^0 +c.c.$ from the analysis reported here; the
resulting change in the $\eta_c$ signal yield is 12.8\%.

The uncertainty due to the nonresonant background shape for each
mode has been estimated by changing the polynomial order from two to
three. The systematic uncertainties due to the fitting ranges are
evaluated by changing them from $1.165 - 1.30$ GeV/$c^2$ to $1.165
-1.25$ GeV/$c^2$ ($\Sigma^0$  and $\bar{\Sigma}^0$), from $1.35 -
1.70$ GeV/$c^2$ to $1.38 - 1.67$ GeV/$c^2$ ($\Lambda(1520)$) and
from $2.76 - 3.06$ GeV/$c^2$ to $2.70 - 3.06$ GeV/$c^2$ ($\eta_c$).
The changes in yields for these variations give systematic
uncertainties due to the choices of fitting ranges, as shown in
Table~\ref{tab:error}.

\begin{table*}[htbp]
\footnotesize
 \centering \caption{Summary of systematic errors for the
branching fraction measurements ($\%$).}
  \setlength{\extrarowheight}{1.0ex}
\begin{tabular}{c|c|c|c|c}
  \hline\hline
                                                         &$J/\psi\rightarrow\bar{\Lambda}\Sigma^{0}$ &$J/\psi\rightarrow\Lambda\bar{\Sigma}^{0}$ & $J/\psi\rightarrow\Lambda\bar{\Lambda}(1520)+c.c.\rightarrow\gamma\Lambda\bar{\Lambda}$&$J/\psi\rightarrow\gamma\eta_{c}\rightarrow\gamma\Lambda\bar{\Lambda}$\\
\hline
   Photon detection                                      &   1                                       &1                                          &1                                                                      &1\\
   Tracking                                        &   4                                       &4                                          &4                                                                      &4\\
   $\Lambda$ and $\bar{\Lambda}$ vertex fits              &   2                                       &2                                          &2                                                                      &2\\
   4C kinematic fit                                      &   2.3                                     &2.3                                        &2.3                                                                    &2.3\\
   Signal shape                                          &   1.3                                     &2.6                                        &4.8                                                                    &7.6\\
   Fitting range                                             &   1.6                                     &0.9                                        &1.4                                                                    &1.4\\
   $\alpha$                                              &   5.5                                    &5.1                                        &10.2                                                                   & - \\
   Fixed  backgrounds                                      &   0.6                                     &0.4                                        &-                                                                      &12.8\\
   Nonresonant background shape                         &   0.3                                     &0.1                                        &1.9                                                                    &1.7\\
   QED correction factor                                 &   0.1                                       &0.1                                        &-
   & - \\
   Cited branching fractions                              &   0.8                                     &0.8                                        &0.8                                                                    &0.8\\
   Number of $J/\psi$                                    &   1.3                                     &1.3                                        &1.3                                                                    &1.3\\\hline
   Total systematic uncertainty                          &   8.0                                    &7.9                                       &12.6                                                                   &16.0\\

  \hline\hline
\end{tabular}
 \label{tab:error}
\end{table*}

The electromagnetic  cross sections for
$\Lambda\bar{\Sigma}^{0}+c.c.$ production through  direct one-photon
exchange and $J/\psi$ decay in $e^+e^-$ can be inferred using the
factorization hypothesis to be~\cite{qederror}
\begin{equation}
\begin{split}
\frac{\sigma(e^+e^-\to \gamma^* \to
\Lambda\bar{\Sigma}^{0})}{\sigma(e^+e^-\to J/\psi\to
\Lambda\bar{\Sigma}^{0})} \approx \frac{\sigma(e^+e^-\to\gamma^*\to
\mu^+\mu^-)}{\sigma(e^+e^-\to J/\psi\to \mu^+\mu^-)}.
\end{split}
\end{equation}
Neglecting interference between $e^+e^-\to\gamma^*\to \mu^+\mu^-$
and $e^+e^-\to J/\psi\to \mu^+\mu^-$, one can obtain, at $\sqrt{s} =
3.097 $ GeV,  $\sigma(e^+e^-\to J/\psi\to
\mu^+\mu^-)=\mathcal{B}(J/\psi \to \mu^+\mu^-)\times
\frac{N_{J/\psi}}{\mathcal{L}}=168\pm 3.2$ nb, where $N_{J/\psi}$
and $\mathcal{L}$ are the number of total $J/\psi$ events ($225.2\pm
2.8\times 10^6$) and the corresponding integrated luminosity
($79631\pm 70 \text{(stat.)}\pm796 \text{(syst.))}$ nb~\cite{number},
respectively. At $\sqrt{s}=3.097$~GeV, $\sigma_{\rm
Born}(e^+e^-\to\gamma^*\to \mu^+\mu^-)$ is 9.05 nb.  From this we
estimate the relative ratio of the QED background from $e^+e^-\to
\gamma^*\to \Lambda\bar{\Sigma}^{0}+c.c.$ to be $(5.4\pm0.1)\%$ of
our measured yield of $J/\psi\to \Lambda\bar{\Sigma}^{0}+c.c.$
events. Therefore, we adjust our result be a factor of 0.946 when we
determine the $J/\psi\to \bar{\Lambda}\Sigma^{0}+c.c.$ branching
fraction value; we use 0.1\% as a systematic error due to the
uncertainty in this correction factor.

\begin{figure}[htbp]
\begin{center}
\includegraphics[width=4.2cm]{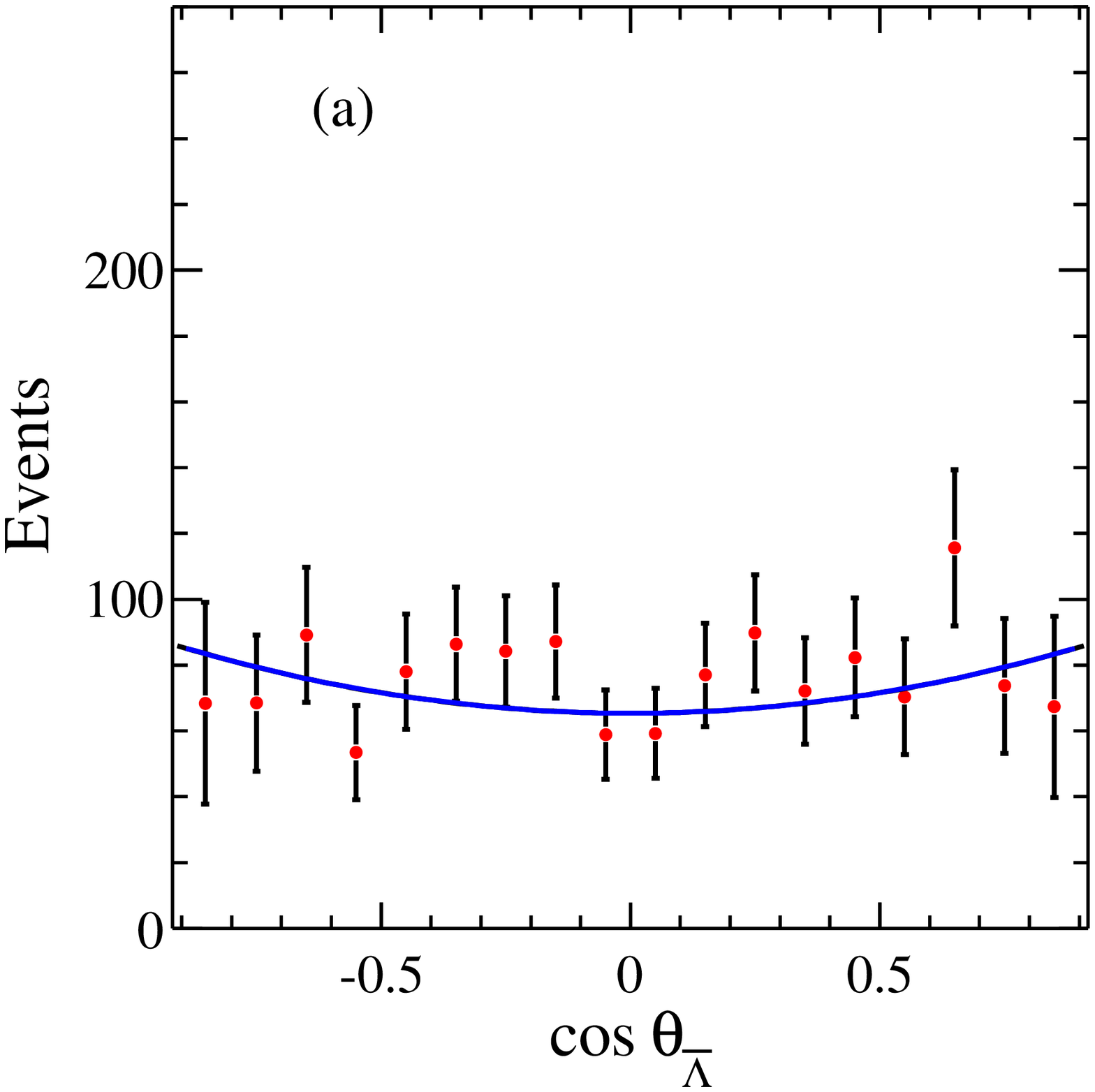}
\includegraphics[width=4.2cm]{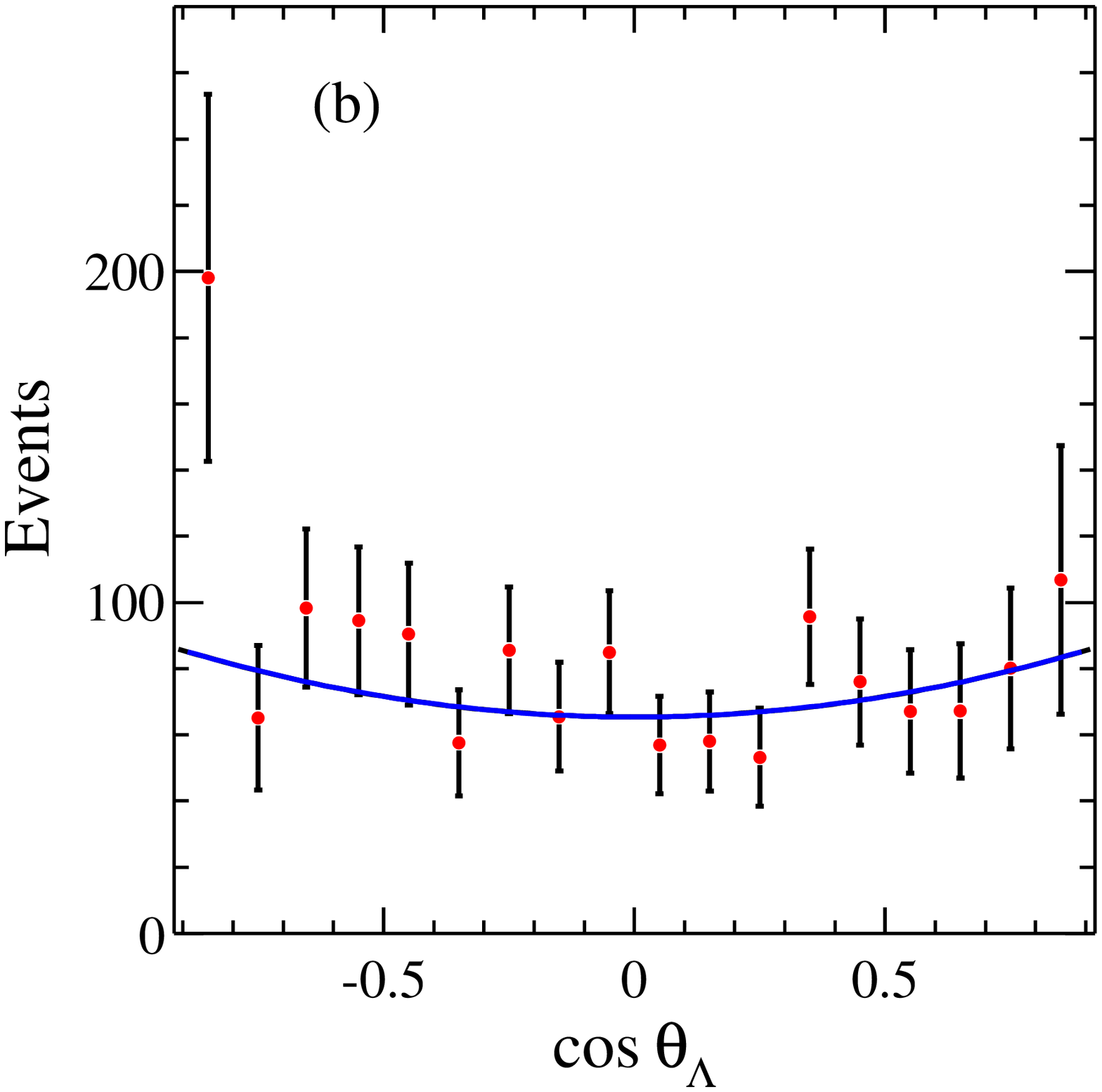}
\caption{ The corrected distributions of $\cos\theta$ for
$\bar\Lambda$ from $J/\psi\to \bar\Lambda\Sigma^{0}$ decay (a),  for
$\Lambda$ from $J/\psi\to \Lambda\bar\Sigma^{0}$ decay (b). The
curves in (a) and (b) present the fits to the function $1+ \alpha
\cos^2 \theta$. The goodness of the fits are $\chi^2/n.d.f =
21/18=1.2$ for $\bar\Lambda$ and $\chi^2/n.d.f = 29/18=1.6$ for
$\Lambda$.}
 \label{anglefit_LS}
\end{center}
\end{figure}

The angular distribution of the baryon in $J/\psi \to B_8 \bar{B}_8$
 decay is expected to have a $1+\alpha \cos^2 \theta$ behavior.
Figures~\ref{anglefit_LS} (a) and (b) show the distributions of
$\cos \theta$ for $\bar\Lambda$ ($J/\psi\to \bar\Lambda\Sigma^{0}$)
and $\Lambda$ ($J/\psi\to \Lambda\bar\Sigma^{0}$), respectively,
after correcting the signal yields for the detection efficiency. A
simultaneous fit to the angular distributions for $\bar\Lambda$  and
$\Lambda$  returns the value  $\alpha~=~0.38\pm 0.39$. The detection
efficiencies are determined with  MC simulation for $J/\psi\to
\Lambda \bar{\Sigma}^0 +c.c.$  using $\alpha=0.38$ in the signal MC
generator.  To estimate the uncertainty originating from the
parameter $\alpha$, we generate MC samples for $\alpha =0.38$ and
for other values in the range  $0.0\sim0.77$. The maximum difference
 is 5.1\% (5.5\%) for
$J/\psi\to \Lambda \bar{\Sigma}^0$ ($\bar\Lambda \Sigma^0$) and is
taken as a systematic error. For $J/\psi\to
\Lambda\bar{\Lambda}(1520) +c.c.$ decay, the detection efficiency is
obtained with  a phase-space MC simulation. We generate MC samples
for $\alpha =0$ and $\alpha=1$ to estimate the uncertainty due to
the unknown parameter $\alpha$. The difference of efficiency of
10.2\% is taken as systematic error for the $J/\psi\to \Lambda
\bar{\Lambda}(1520) +c.c.$.

The branching fraction for the $\Lambda\rightarrow p\pi$ decay is
taken from the PDG~\cite{pdg}; the 0.8\% uncertainty  is taken as a
systematic uncertainty in our measurements. The uncertainty in the
number of $J/\psi$ decays in our data sample is 1.3\%~\cite{number}.
The total systematic uncertainties for the branching fraction
measurements are obtained by adding up the contributions from all
the systematic sources in quadrature as summarized in
Table~\ref{tab:error}.

\section{Results and Discussion}

The branching fractions are calculated with ${\mathcal B} =
N_S/(N_{J/\psi} \epsilon {\mathcal B}_{p\pi}^2)$, where $N_S$ and
$\epsilon$ are the number of signal events and the detection
efficiency, listed in Table~\ref{sum:eff_yield}. Here $N_{J/\psi} =
(225.2\pm 2.8)\times 10^6$~\cite{number} is the number of $J/\psi$
events, and ${\mathcal B}_{p\pi}$ is the branching fraction of the
$\Lambda \to p\pi$ taken from the PDG~\cite{pdg}. The calculated
branching fractions, along with the PDG~\cite{pdg} limits, are
listed in Table~\ref{tab:sum}.
\begin{table}[htbp]
\centering \caption{Branching fractions ($10^{-5}$) from this
analysis, where the first errors are statistical and the second ones
are systematic, and the PDG values~\cite{pdg} for comparison. The
upper limits are at the 90\% C.L.. }
  \setlength{\extrarowheight}{1.0ex}
\begin{tabular}{c|c|c}
  \hline\hline
   $J/\psi$ decay mode                                                                            &BESIII                           &PDG\\
\hline
   $\bar{\Lambda}\Sigma^{0}$                                            &$1.46\pm0.11\pm0.12$    &$<7.5$\\
   $\Lambda\bar{\Sigma}^{0}$                                            &$1.37\pm0.12\pm0.11$    &$<7.5$\\
   $\gamma\eta_{c}(\eta_c\to\Lambda\bar{\Lambda})$                      &$1.98\pm0.21\pm0.32$    & -\\
   $\Lambda\bar{\Lambda}(1520)+c.c. (\bar{\Lambda}(1520)\to \gamma\bar{\Lambda})$    &$<0.41$  &-\\
\hline\hline
\end{tabular}
 \label{tab:sum}
\end{table}
\begin{table}[htbp]
\footnotesize
 \centering \caption{Amplitude parametrizations from
~\cite{kopke,kowalski,dhwei} for $J/\psi$ decay to a pair of octet
baryons. General expressions in terms of a singlet \emph{A}, as well
as symmetric and antisymmetric charge-breaking (\emph{D}, \emph{F})
and mass-breaking terms (\emph{D}$^{'}$, \emph{F}$^{'}$) are given.
Here $\delta$ is the relative phase between one-photon and
gluon-mediated hadronic decay amplitudes. Except for the branching
fraction for $J/\psi\rightarrow \Lambda\bar{\Sigma}^{0}+c.c.$ decay
(marked with an asterisk) from this measurement and for $J/\psi\to
p\bar{p}$, $n\bar{n}$ from the recent BESIII
measurements~\cite{bianjm}, the other branching fractions (${\cal
B}$) are taken from the PDG~\cite{pdg}. }
\begin{tabular}{c|c|c}
  \hline\hline
  Decay mode & Amplitude  &
  $\mathcal{B}$($\times10^{-3}$)\\\hline
  $p \bar{p}$ & $A+e^{i\delta}(D+F)+D^{'}+F^{'}$ & ($2.112\pm0.031)$~\cite{bianjm}\\
  $n \bar{n}$ & $A-e^{i\delta}(2D)+D^{'}+F^{'}$  & ($2.07\pm0.17)$~\cite{bianjm}\\
  $\Sigma^{+}\bar{\Sigma}^{-}$ & $A+e^{i\delta}(D+F)-2D^{'}$ & ($1.50\pm0.24$)\\
  $\Sigma^{0}\bar{\Sigma}^{0}$ & $A+e^{i\delta}(D)-2D^{'}$ & ($1.29\pm0.09$)\\
  $\Xi^{0}\bar{\Xi}^{0}$ & $A-e^{i\delta}(2D)+D^{'}-F^{'}$ & ($1.20\pm0.24$)\\
  $\Xi^{-}\bar{\Xi}^{+}$ & $A+e^{i\delta}(D-F)+D^{'}-F^{'}$ & ($0.85\pm0.16$)\\
  $\Lambda\bar{\Lambda}$ & $A-e^{i\delta}(D)+2D^{'}$ &  ($1.61\pm0.15$)\\
  $\bar{\Lambda}\Sigma^{0}(\Lambda\bar{\Sigma}^{0})$ & $(\sqrt{3}D)$ &$(0.014\pm0.002)^*$\\
  \hline\hline
\end{tabular}
 \label{tab:fit_amp}
\end{table}
\begin{table*}[htbp]
\centering \caption{Constraint fit results for the amplitude
parametrizations in terms of a singlet \emph{A}, symmetric and
antisymmetric charge-breaking (\emph{D}, \emph{F}), mass-breaking
(\emph{D}$^\prime$, \emph{F}$^\prime$) terms and a relative phase
$\delta$ as listed in Table~\ref{tab:fit_amp}. The fit is
constrained to the measured branching fractions from PDG~\cite{pdg}
and Ref.~\cite{bianjm}, as listed in Table~\ref{tab:fit_amp}, as
well as the measurement in this analysis. The $\chi^2$/d.o.f. is
$1.01/2.0$ for the fit.  Similar fitting results from
Ref.~\cite{dhwei} are also shown for comparison.}
\begin{tabular}{c|c|c|c|c|c|c}
  \hline\hline
                      &      $A$           &         $D$         &        $F$         &        $D^\prime$        &        $F^\prime$          &      $\delta$       \\
\hline
   our fit            & $1.000\pm 0.044$ & $-0.058\pm 0.005$ & $0.231\pm 0.140$ & $0.015\pm 0.028$ & $-0.027\pm 0.045$  &  $(76\pm 11)^\circ$ \\
   Ref.~\cite{dhwei}  & $1.000\pm 0.028$ &   0 (fixed)       & $0.341\pm 0.085$ & $0.032\pm 0.041$ & $-0.050\pm 0.070$  & $(106\pm 8)^\circ$  \\
  \hline\hline
\end{tabular}
\label{tab:fitzhf}
\end{table*}

Our measurement of the branching fraction for $J/\psi\rightarrow
\Lambda\bar{\Sigma}^{0}+c.c.$ decay can shed light on the $SU(3)$
breaking mechanism. The amplitude for $J/\psi$ decay to a pair of
octet baryons can be parametrized in terms of a $SU(3)$ singlet
\emph{A}, as well as symmetric and antisymmetric charge-breaking
(\emph{D}, \emph{F}) and mass-breaking (\emph{D}$^{'}$,
\emph{F}$^{'}$) terms, as described in
Refs.~\cite{kopke,kowalski,dhwei} and listed in
Table~\ref{tab:fit_amp}, where $\delta$ is used to designate the
relative phase
 between the one-photon and gluon-mediated hadronic decay amplitudes.
According to these amplitude parametrizations the
$J/\psi\rightarrow\bar{\Lambda}\Sigma^{0}+c.c.$ branching fraction
measurement is important for the determination of the symmetric
charge-breaking term \emph{D}.  In Ref.~\cite{dhwei}, a constrained
fit to the measured branching fractions of $J/\psi \to B_8\bar{B}_8$
is performed to extract the values of the parameters $A$, $F$,
$D^\prime$, $F^\prime$ and $\delta$ using the
Table~\ref{tab:fit_amp} amplitude parametrizations. In the previous
fit~\cite{dhwei}, $D=0$ was assumed, {\it i.e.,} ${\mathcal
B}(J/\psi\rightarrow\Lambda\bar{\Sigma}^{0}+c.c.)=0$. We perform
another fit that includes our new measurement and includes a nonzero
value for $D$. The fit results are listed in Table~\ref{tab:fitzhf}.
In comparison to the Ref.~\cite{dhwei} results, the value for the
relative phase $\delta$ has changed significantly,  while the $A$,
$D^\prime$, $F$ and $F^\prime$ values do not change significantly.
The measurement of the isospin violating decay  $J/\psi \to
\bar\Lambda \Sigma^0+c.c.$ also provides useful information on  the
mechanisms for $J/\psi\to B_8\bar B_{10}$ decays, where the large
$A$-term is absent~\cite{kopke,kowalski}.

\section{summary}

In summary, with a sample of $(225.2\pm 2.8)\times 10^6$ $J/\psi$
events in the BESIII detector, the
$J/\psi\rightarrow\gamma\Lambda\bar{\Lambda}$ decay has been
studied. The branching fractions of
$J/\psi\rightarrow\bar{\Lambda}\Sigma^{0}$,
$J/\psi\rightarrow\Lambda\bar{\Sigma}^{0}$ and
$J/\psi\rightarrow\gamma\eta_{c}(\eta_c\to\Lambda\bar{\Lambda})$ are
measured for the first time as: ${\cal
B}(J/\psi\rightarrow\bar{\Lambda}\Sigma^{0}) = (1.46\pm0.11\pm0.12)
\times10^{-5}$, ${\cal B}(J/\psi\rightarrow\Lambda\bar{\Sigma^{0}}) =
(1.37\pm0.12\pm0.11) \times10^{-5}$ and ${\cal
B}(J/\psi\rightarrow\gamma\eta_{c})\times {\cal
B}(\eta_{c}\rightarrow\Lambda\bar{\Lambda}) =
(1.98\pm0.21\pm0.32)\times10^{-5}$, respectively, where the
uncertainties are statistical and systematic.  Using the PDG
value~\cite{pdg} for $J/\psi \to \gamma \eta_c$, we obtain ${\cal
B}(\eta_c\to \Lambda \bar{\Lambda}) =
(1.16\pm0.12\pm0.19\pm0.28\mbox{~(PDG)})\times10^{-3}$, where the
third error is from the error on ${\cal B}(J/\psi\rightarrow \gamma
\eta_c)$. Using $B^{\pm} \to \Lambda \bar\Lambda K^{\pm}$ decay the
Belle experiment measured ${\cal B}(\eta_c\to \Lambda
\bar{\Lambda})=(0.87^{+0.24+0.09}_{-0.21-0.14}\pm0.27\mbox{~(PDG)})\times10^{-3}$~\cite{belle},
which is consistent with our result within error. No evidence for the
decay of $J/\psi\rightarrow\Lambda\bar{\Lambda}(1520)+c.c.$ is found,
and an upper limit for the branching fraction is determined to be
${\cal B}(J/\psi\rightarrow\Lambda\bar{\Lambda}(1520)+c.c.)\times
{\cal B}(\Lambda(1520)\rightarrow\gamma\Lambda)<4.1 \times10^{-6}$ at
the 90\% confidence level. Results are listed in Table~\ref{tab:sum}
and compared with previous measurements.

\begin{acknowledgements}
The BESIII collaboration thanks the staff of BEPCII and the
computing center for their hard efforts. This work is supported in
part by the Ministry of Science and Technology of China under
Contract No. 2009CB825200; National Natural Science Foundation of
China (NSFC) under Contracts Nos. 10625524, 10821063, 10825524,
10835001, 10935007, 11125525; Joint Funds of the National Natural
Science Foundation of China under Contracts Nos. 11079008, 11179007;
the Chinese Academy of Sciences (CAS) Large-Scale Scientific
Facility Program; CAS under Contracts Nos. KJCX2-YW-N29,
KJCX2-YW-N45; 100 Talents Program of CAS; Istituto Nazionale di
Fisica Nucleare, Italy; Ministry of Development of Turkey under
Contract No. DPT2006K-120470; U. S. Department of Energy under
Contracts Nos. DE-FG02-04ER41291, DE-FG02-91ER40682,
DE-FG02-94ER40823; U.S. National Science Foundation; University of
Groningen (RuG) and the Helmholtzzentrum fuer Schwerionenforschung
GmbH (GSI), Darmstadt; WCU Program of National Research Foundation
of Korea under Contract No. R32-2008-000-10155-0.
\end{acknowledgements}



\begin{thebibliography}{11}
%
\bibitem{Tab1}
D. M. Asner {\it et al.}, Int. J. Mod. Phys. A {\bf 24}, 499 (2009).

%
\bibitem{kopke}
L. Kopke and N. Wermes, Phys. Rept. {\bf 174}, 67 (1989).

%
\bibitem{kowalski}
H. Kowalski and T. F. Walsh, Phys. Rev. D {\bf 14}, 852 (1976).

%
\bibitem{mark1}
I. Peruzzi {\it et al.} (MARK I Collaboration), Phys. Rev. D {\bf
17}, 2901 (1978).

%
\bibitem{dulat}
  S. Dulat, J. J. Wu and B. S. Zou,
  Phys. Rev. D {\bf 83}, 094032 (2011).

%
\bibitem{kaxiras}
E. Kaxiras, E. J. Moniz and M. Soyeur, Phys. Rev. D {\bf 32}, 695
(1985).

%
\bibitem{darewych}
J. W. Darewych, M. Horbatsch and R. Koniuk, Phys. Rev. D {\bf 28},
1125 (1983).

%
\bibitem{warns}
M. Warns, W. Pfeil and H. Rollnik, Phys. Lett. B {\bf 258}, 431
(1991).

%
\bibitem{umino}
Y. Umino and F. Myhrer,  Nucl. Phys. A {\bf 529}, 713 (1991); Nucl.
Phys. A {\bf 554}, 593 (1993).

%
\bibitem{bijker}
R. Bijker, F. Iachello and A. Leviatan, Annals Phys. {\bf 284}, 89
(2000).

%
\bibitem{LangYu}
L. Yu, X. L. Chen, W. Z. Deng and S. L. Zhu, Phys. Rev. {\bf D} 73,
114001 (2006).

%
\bibitem{mast}
T. S. Mast, M. Alston-Garnjost, R. O. Bangerter, A.
Barbaro-Galtieri, L. K. Gershwin, F. T. Solmitz, and R. D. Tripp,
Phys. Rev. Lett. {\bf 21}, 1715 (1968).

%
\bibitem{bertini}
R. Bertini, Nucl. Phys. B {\bf 279}, 49 (1987); R. Bertini {\it et
al.}, SACLAY-DPh-N-2372 (unpublished).

%
\bibitem{antipov}
Y. M. Antipov {\it et al.} (SPHINX Collaboration), Phys. Lett. B
{\bf 604}, 22 (2004).

%
\bibitem{taylor}
S. Taylor {\it et al.} (CLAS Collaboration), Phys. Rev. C {\bf 71},
054609 (2005).


%
\bibitem{belle} C.~H.~Wu {\it et al.} (Belle Collaboration), Phys. Rev. Lett.
 {\bf 97}, 162003 (2006).

%

%

%
\bibitem{number} M. Ablikim {\it et al.} (BES Collaboration), Phys. Rev. D
{\bf 83}, 012003 (2011).

%
\bibitem{dect8}
  M. Ablikim {\it et al.} (BES Collaboration), Nucl. Instrum. Meth. A {\bf 614} 345, (2010).

%
\bibitem{geant1} S. Agostinelli {\it et al.} (GEANT4 Collaboration), Nucl. Instrum. Methods
Phys. Res., Sect. A {\bf 506}, 250 (2003).

%
\bibitem{geant2} J. Allison {\it et al.}, IEEE Trans. Nucl. Sci. {\bf 53}, 270 (2006).


\bibitem{evtgen} D.~J.~Lange, Nucl. Instrum. Meth. A {\bf 462}, 152 (2001).

\bibitem{kkmc} S.~Jadach, B.~F.~L. Ward and Z.~Was, Comput. Phys. Commun. {\bf 130}, 260 (2000);
               S.~Jadach, B.~F.~L. Ward and Z.~Was \jprd{\bf 63}, 113009 (2001).

\bibitem{pdg}
The Review of Particle Physics, C. Amsler {\it et al.}, J. Phys. G
{\bf 37}, 075021 (2010).
\bibitem{lund} J. C. Chen, G. S. Huang, X. R. Qi, D. H. Zhang, and Y. S. Zhu, Phys.
Rev. D {\bf 62}, 034003 (2000).


%
\bibitem{ppbar}  M. Ablikim {\it et al.} (BES Collaboration),
 Phys. Rev. Lett. {\bf 108}, 112003 (2012).

%
\bibitem{anashin}
V.~V.~Anashin {\it et al.} (KEDR Collaboration), arXiv:1012.1694
[hep-ex]

%
\bibitem{cleoc} R.~E.~Mitchell {\it et al.} (CLEO Collaboration),
Phys. Rev. Lett. {\bf 102}, 011801 (2009).

%
\bibitem{bes3etac} M. Ablikim {\it et al.} (BES Collaboration), Phys. Rev. Lett. {\bf 108},
222002 (2012).
%

%
\bibitem{etacfit}
We also considered possible interference effects between $\eta_{c}$
signal and non-resonance backgrounds. With the assumption of all the
non-resonant backgrounds are from $0^{-+}$ phase space, we obtained
two solutions:  $\phi = 4.74 \pm
0.29(\text{stat.})$~rad~(constructive) or $\phi = 1.46 \pm
0.23(\text{stat.})$~rad~(destructive), where $\phi$ is the relative
 phase between $\eta_c$ resonance and non-resonance amplitudes.
 The constructive (destructive) interference results in ${\mathcal
B}(J/\psi\to \gamma\eta_{c}\to \gamma\Lambda\bar{\Lambda})=(1.36 \pm
0.31(\text{stat.}))\times10^{-5}$ ($(3.48 \pm
0.70(\text{stat.}))\times10^{-5}$). The fit method is similar to
that described in Ref.~\cite{bes3etac}.
%
\bibitem{qederror}
J. G. Korner, M. Kuroda, Phys. Rev. D {\bf 16}, 2165 (1977).


\bibitem{dhwei}
D. H. Wei, J. Phys. G {\bf 36}, 115006 (2009).

\bibitem{bianjm}
M. Ablikim {\it et al.} (BES Collaboration), arXiv:1205.1036
[hep-ex].


\end{thebibliography}
\end{document}